\documentclass[a4paper]{jpconf}
\usepackage{graphicx}
\usepackage{lineno}
\newcommand {\pT}{\ensuremath{p_{\mathrm{T}}}}
\newcommand {\meanpT}{$\langle \pT \rangle$}

\newcommand {\dnchdeta}{d$N_{\mathrm{ch}}$/d$\eta$}
\newcommand {\meandnchdeta}{$\langle$d$N_{\mathrm{ch}}$/d$\eta$$\rangle$}
\newcommand {\meandnchdetaeta}{$\langle$d$N_{\mathrm{ch}}$/d$\eta$$\rangle_{\left | \eta \right | < 0.5}$}

\newcommand {\AuAu}{\mbox{Au--Au}}
\newcommand {\XeXe}{\mbox{Xe--Xe}}
\newcommand {\PbPb}{\mbox{Pb--Pb}}
\newcommand {\pPb}{\mbox{p--Pb}}
\newcommand {\dAu}{\mbox{d--Au}}

\newcommand {\s}{$\sqrt{s}$}
\newcommand {\sNN}{$\sqrt{s_{\mathrm{NN}}}$}

\newcommand {\gmom} {GeV/$c$}

\newcommand {\fmc} {fm/$c$}
\newcommand{\K}{\mbox{$\mathrm {K}$}}

\newcommand{\pion}{\mbox {$\mathrm {\pi}$}}

\newcommand{\rmLambda}{\mbox{$\mathrm {\Lambda}$}}

\newcommand{\rmSigma}{\mbox{$\mathrm {\Sigma}$}}
\newcommand{\rmXi}{\mbox{$\mathrm {\Xi}$}}

\newcommand{\kstarZ}{\mbox{\K$^{*}$(892)$^{\mathrm{0}}$}}
\newcommand{\kstarch}{\mbox{\K$^{*}$(892)$^{\mathrm{\pm}}$}}
\newcommand{\simplekstarch}{\mbox{\K$^{*\mathrm{\pm}}$}}
\newcommand{\simplekstarZ}{\mbox{\K$^{*\mathrm{0}}$}}
\newcommand{\phir}{\mbox{$\mathrm {\phi}$(1020)}}
\newcommand{\rmphi}{\mbox{$\mathrm{\phi}$}}

\newcommand{\kstar}{\mbox{\K$^{*}$}}
\newcommand{\rmrho}{\mbox{$\mathrm {\rho}$(770)$^{\mathrm{0}}$}}
\newcommand{\simplerho}{\mbox{$\mathrm{\rho}^{\mathrm{0}}$}}
\newcommand{\Sigmastar}{\mbox{$\mathrm{\Sigma}$(1385)$^{\mathrm{\pm}}$}}

\newcommand{\simplesigmastar}{\mbox{$\mathrm{\Sigma}^{\mathrm{*\pm}}$}}

\newcommand{\simplelambdastar}{\mbox{$\mathrm{\Lambda}^{\mathrm{*}}$}}
\newcommand{\simplexistar}{\mbox{$\mathrm{\Xi}^{\mathrm{*}}$}}
\newcommand{\Lambdastar}{\mbox{$\mathrm{\Lambda}$(1520)$^{\mathrm{0}}$}}
\newcommand{\Xistar}{\mbox{$\mathrm {\Xi}$(1530)$^{\mathrm{0}}$}}

\begin{document}
\title{Energy and multiplicity dependence of hadronic resonance production with ALICE at the LHC}

\author{Angela Badal\`a for the ALICE Collaboration}

\address{INFN - Sezione di Catania , Via S. Sofia 64, 95123, Catania (Italy)}

\ead{angela.badala@ct.infn.it}

\begin{abstract}
The study of hadronic resonances plays an important role both in pp and in heavy-ion collisions. Since the lifetimes of short-lived resonances are comparable with the lifetime of the fireball formed in heavy-ion collisions, regeneration and re-scattering effects can modify the measured yields, especially at low transverse momentum. Measurements in pp collisions at different energies constitute a baseline for studies in heavy-ion collisions and provide constraints for tuning QCD-inspired event generators. Furthermore, high multiplicity pp collisions, where the density and the volume of the system are expected to be larger compared to minimum bias pp collisions, can help in the search for the onset of collective phenomena. Here we present recent results on short-lived hadronic resonances obtained by the ALICE experiment at LHC energies in different collision systems (pp, \pPb~and \PbPb)  including new results obtained in \XeXe~collisions. The ALICE results on transverse momentum spectra, yields and their ratios to long-lived particles will be discussed.
\end{abstract}

\section{Introduction}

One of the main motivations for studying hadronic resonance production in heavy-ion collisions is to learn more about the properties (temperature and lifetime) of the hadronic phase, the late-stage evolution of these collisions.  The decay products of short-lived resonances may re-scatter in the hadronic phase,  leading  to  a  reduction  in  the  measurable  resonance  yields;  conversely, resonance  yields  may also be regenerated by pseudo-elastic scattering of hadrons through a resonance 
state~\cite{Bleicher,Torrieri,Markert,Vogel}.
%[38] M. Bleicher and H. St ̈ocker, “Dynamics and freeze-out of hadron resonances at RHIC,”J. Phys. G46730(2004) S111–S118,arXiv:hep-ph/0312278.468
%[39]  G. Torrieri and J. Rafelski, “Strange hadron resonances as a signature of freeze-out dynamics,” Phys. Lett. B509(2001) 239–245,arXiv:hep-ph/0103149.470
%[40]  C. Markert, R. Bellwied, and I. Vitev, “Formation and decay of hadronic resonances in the QGP,” Phys. Lett. B669(2008) 92–97,arXiv:0807.1509.472
%[41]  S. Vogel, J. Aichelin, and M. Bleicher, “Resonances as a possible observable of hot and dense nuclear matter,”J. Phys. G37(2010) 094046,arXiv:1001.3260 
%[42] M. Bleicher and J. Aichelin, “Strange resonance production: probing chemical and thermal freeze-out in relativistic heavy-ion collisions,”Phys. Lett. B530(2002) 81–87,476arXiv:hep-ph/0201123.477
%[43]  A. G. Knospe et al., “Hadronic resonance production and interaction in partonic and hadronic matter in EPOS3 with and without the hadronic afterburner UrQMD,”Phys. Rev. C93(2016)479014911,arXiv:1509.07895
Moreover, hadronic resonances, along with stable hadrons, allow the study of properties of heavy-ion collisions, both in the early (quark-gluon plasma) and late (hadronic) stages of their evolution. 
%Energy loss of partons in nuclear matter can be studied by measuring the nuclear modification factors \RAA~and \RpPb. 
The various mechanisms that may determine the shapes of particle \pT~spectra, including parton fragmentation, quark  recombination, hydrodynamic flow, re-scattering,  and  regeneration,  can be studied through comparison of different measurements of multiple particle species with differing masses  and quark content. 
  
Particle production in pp collisions at collider energies originates from the interplay of perturbative (hard) and non-perturbative (soft) QCD processes, which can only be modeled phenomenologically. The measurements in pp collisions at different energies constitute a baseline for studies in heavy-ion collisions and provide constraints for tuning  event generators such as PYTHIA~\cite{PYTHIA8} and EPOS-LHC~\cite{EPOS_LHC}. Recent observations on the enhancement of (multi-)strange hadrons~\cite{Strangeness_pPb,Strangeness_pp}, 
double-ridge structure~\cite{Long_range_CMS,Long_range_ALICE}, non-zero v$_2$ coefficients~\cite{V2_pp} and mass ordering in hadron \pT~spectra~\cite{Multiplicity_dep_pp7TeV,Multiplicity_dep_pPb} indicate the presence of collective-like phenomena in small systems such as \pPb~and pp collisions at LHC energies. Furthermore, a continuous transition of light-flavour hadron to pion ratios as a function of charged particle multiplicity density \dnchdeta~from pp to \pPb~and then to \PbPb~collisions was observed~\cite{Multiplicity_dep_pp7TeV,Multiplicity_dep_pPb}. However the origin of these effects is not yet fully understood and it remains an open question whether the underlying mechanisms are the same in these three collision systems.

The ALICE experiment has measured the production of a rich set of hadronic resonances, such as \rmrho~\cite{Rho_pbpb}, 
\kstarZ~\cite{phi_kstar_ALICE},
%{phi_kstar_pp,phi_kstar_ppb,phi_kstar_pbpb,phi_kstar_highpt}, 
\phir~\cite{phi_kstar_ALICE},
%{phi_kstar_pp,phi_kstar_ppb,phi_kstar_pbpb,phi_kstar_highpt}, 
\Sigmastar~\cite{Sigmastar_pp,Sigmastar_ppb}, \Lambdastar~\cite{Lambdastar_pbpb}~and \Xistar~\cite{Sigmastar_pp,Sigmastar_ppb} in pp, \pPb~and \PbPb~collisions at various energies at the LHC, and also in \XeXe~collisions at \sNN~=~5.44~TeV collected in late 2017.
This paper presents the recent results on hadronic resonance production from the ALICE experiment. A detailed review of the ALICE detector and its perfomance can be found in ~\cite{alice,alice_perf}.

\begin{figure*}
\centering
%\vskip -0.65cm
\includegraphics[width=15cm]{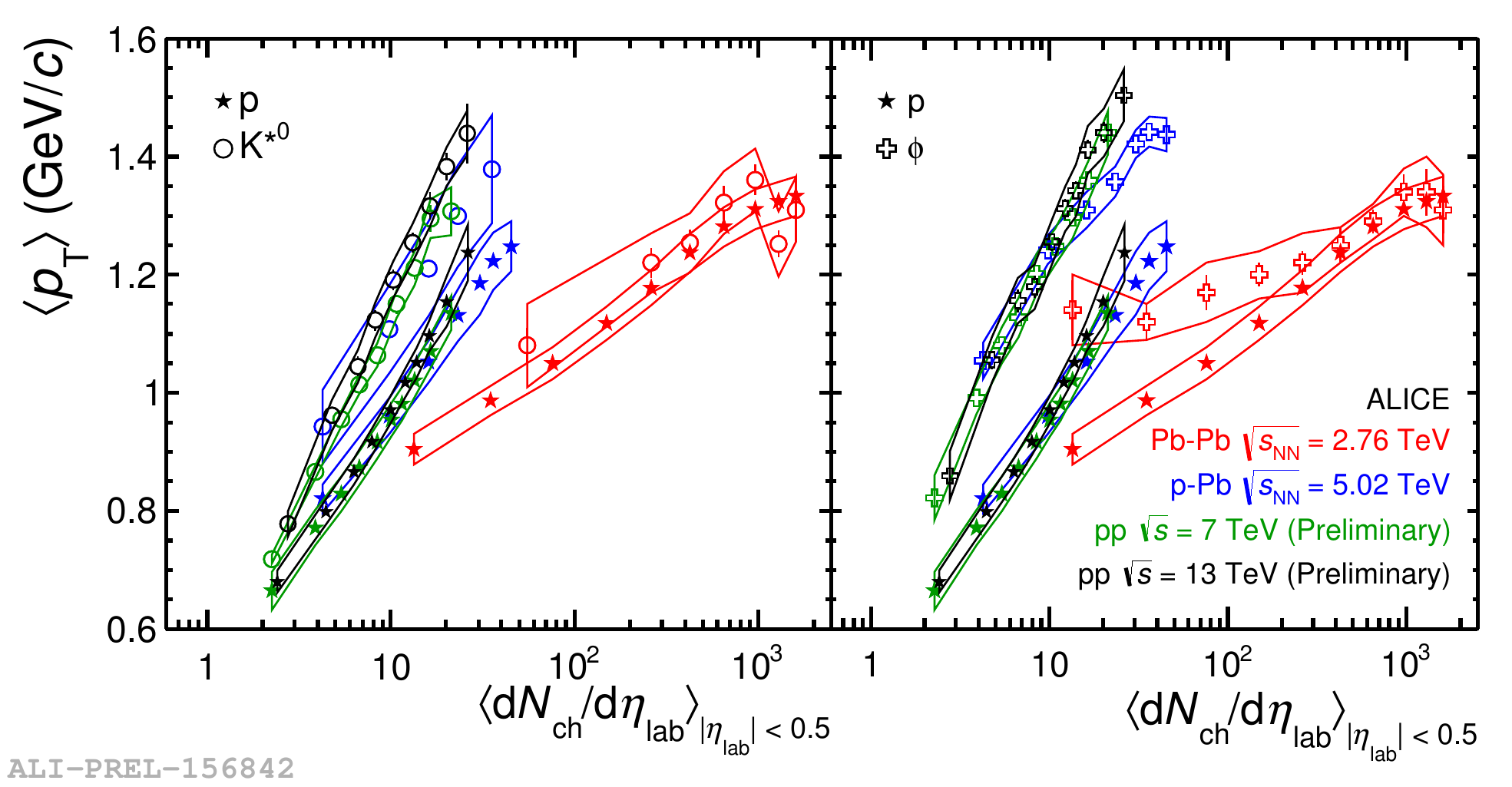}
\caption{Mean transverse momentum \meanpT~values of \simplekstarZ, p and \rmphi~in various collision systems~\cite{phi_kstar_ALICE,Multiplicity_dep_pPb,Multiplicity_dep_PbPb}
%{phi_kstar_pbpb,phi_kstar_ppb,Multiplicity_dep_pPb,Multiplicity_dep_PbPb} 
as a function of charged-particle multiplicity density at mid-rapidity}
\label{fig:meanpT}       % Give a unique label
\end{figure*}

\begin{figure*}
\centering
%\vskip -0.65cm
\includegraphics[width=7.5cm]{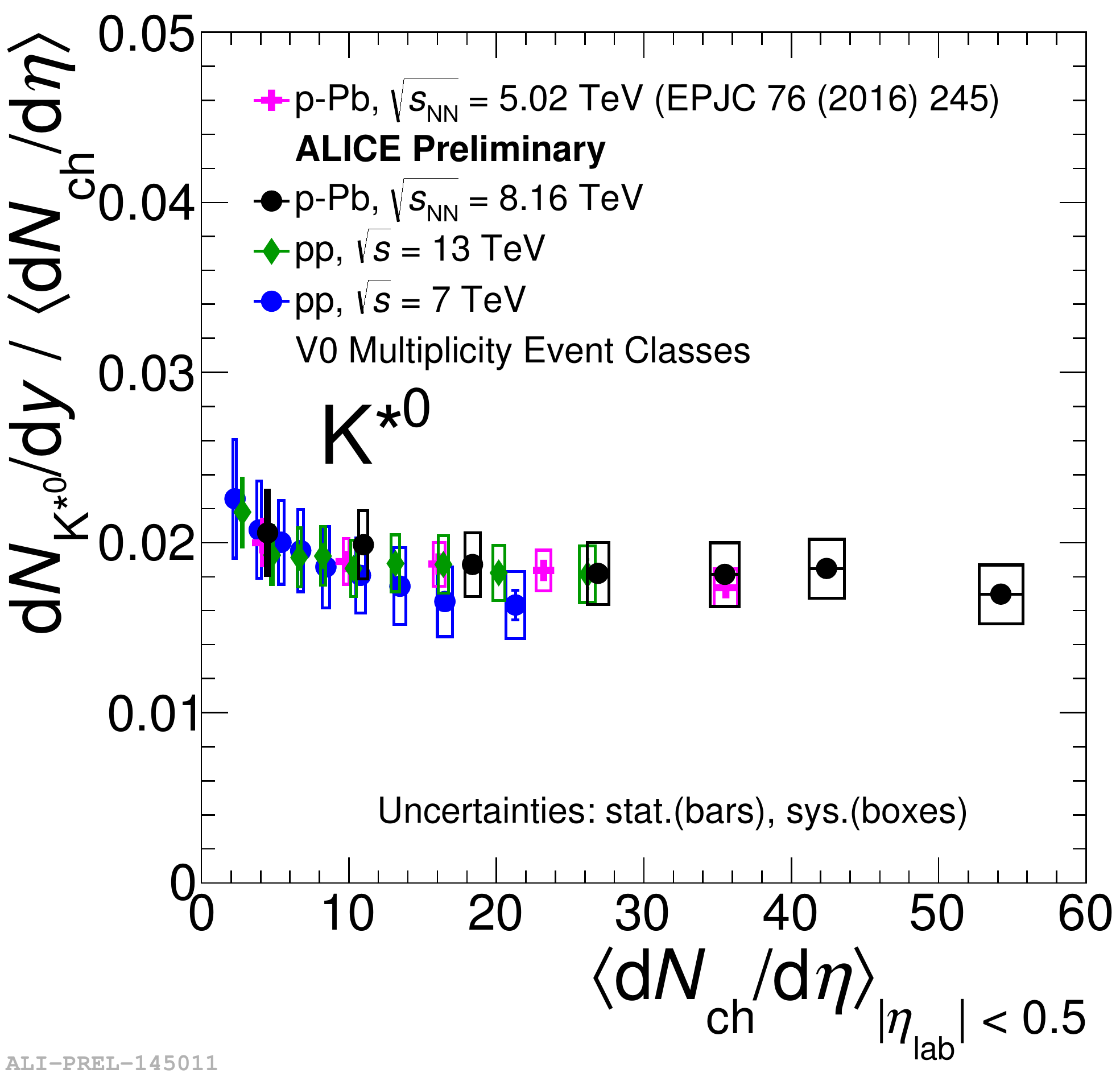}
\includegraphics[width=7.5cm]{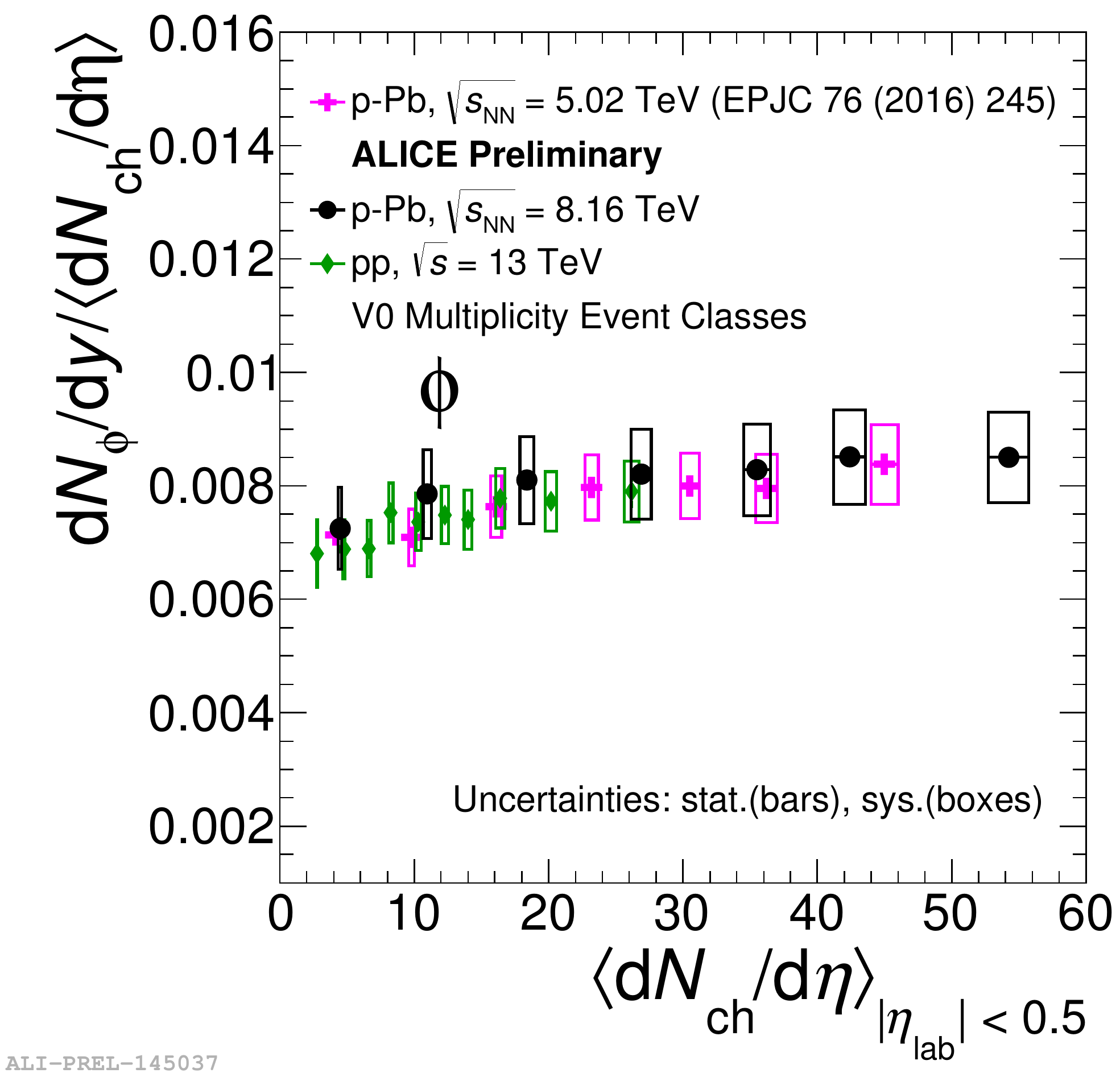}
\caption{The integrated yield normalized to \meandnchdeta~as a function of charged-particle multiplicity density \meandnchdetaeta~for \simplekstarZ~(left panel) and \rmphi~(right panel)}
\label{fig:dNdeta}       % Give a unique label
\end{figure*}

\section{Spectral shape and mean \pT}

In \PbPb~collisions and in elementary collisions \pT~spectra get harder with increasing multiplicity. While in \PbPb~collisions this effect is
attributed to collective expansion, the origin of collective flow-like effect in small collision systems is not clear.  However some collective expansion models as EPOS-LHC~\cite{EPOS_LHC} or models as PYTHIA~\cite{PYTHIA8} or DIPSY~\cite{DIPSY} with the color reconnection mechanism are able to mimic these collective effects in small collision systems.
The mean transverse momentum \meanpT~provides first-order characterization of spectral shapes. The \meanpT~values of the \simplekstarZ, p, and \rmphi~(which all have similar masses) are shown in Fig.~\ref{fig:meanpT} for different collision systems. In central {\mbox{A--A}} collisions, mass ordering of \meanpT~values is observed; particles with similar masses have similar \meanpT. This behavior has been interpreted as evidence that radial flow could be a dominant factor in determining the shapes of hadron \pT~spectra in central
{\mbox{A--A}} collisions. However, this mass ordering breaks down for peripheral \PbPb~collisions, as well as for \pPb~and pp collisions, where the proton is observed to have lower \meanpT~values than the two mesonic resonances. The \meanpT~values in pp and \pPb~collisions also follow different trends and rise faster with multiplicity than in \PbPb~collisions.
The slope of \pT~spectra in selected multiplicity interval, normalized to the corresponding minimum bias spectra, show an increase with increasing multiplicity in the transverse momentum region \pT~$<$~4~\gmom~and have the same slope for higher 
\pT~\cite{Multiplicity_dep_pp7TeV}.  
The multipliciy dependence of the slope of the spectra at low \pT~was also observed for stable hadrons~\cite{Ganoti_spectra}.

It is interesting to note that the yield of \simplekstarZ~and \rmphi, normalized by the mean charge particle multiplicity density \meandnchdetaeta~(Fig.~\ref{fig:dNdeta}), shows a behavior with \meandnchdetaeta~that is independent of collision energy and system for pp and \pPb~collisions.
\begin{figure*}
\centering
%\vskip -0.65cm
\includegraphics[width=13cm]{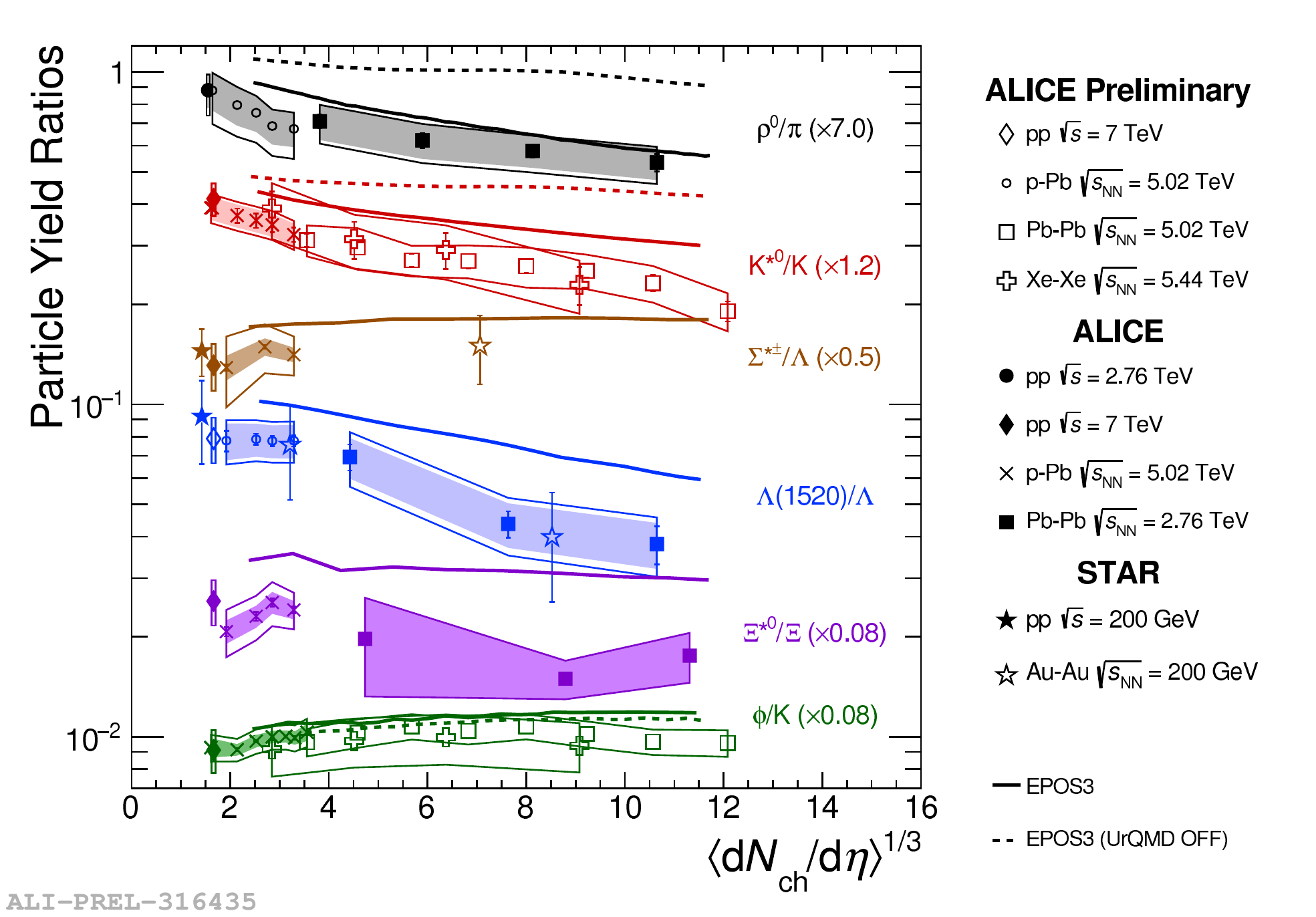}
\caption{Ratios of resonances and ground state yields \simplerho/\pion~\cite{Rho_pbpb}, 
\simplekstarZ/\K~\cite{phi_kstar_ALICE},
%{phi_kstar_pp,phi_kstar_ppb,phi_kstar_pbpb,phi_kstar_highpt}, 
\simplesigmastar/\rmSigma~\cite{Sigmastar_ppb}, \simplelambdastar/\rmLambda~\cite{Lambdastar_pbpb}, \simplexistar/\rmXi~~\cite{Sigmastar_pp,Sigmastar_ppb} and \rmphi/\K~~\cite{phi_kstar_ALICE}
%{phi_kstar_pp,phi_kstar_ppb,phi_kstar_pbpb,phi_kstar_highpt} 
as a function of the cubic root of the charged particle multiplicity density \dnchdeta~for various collision systems. STAR data are
 also shown for \simplesigmastar/\rmSigma~\cite{STAR_Sigmastar}.The error bars show the statistical uncertainty, while the empty and dark-shaded boxes show the total systematic uncertainty and the uncorrelated contribution across multiplicity bins, respectively. Continuous lines show the distributions obtained  by EPOS3 model~\cite{EPOS}. The  dashed lines  represent the distributions estimated by EPOS3 if no final state interactions are taken into account.}
\label{fig:ratios}       % Give a unique label
\end{figure*}

\section{Resonance ratios}

The measured hadronic resonance yields may be influenced by several factors: initial yield at chemical freeze-out, resonance lifetime, scattering cross-section of resonance decay daughters and lifetime of the hadronic phase of the system evolution. A  comparison of measured resonance yields to the production rate of its stable counterpart can provide information about the late stage of the system evolution. Hadronic resonances reported here are reconstructed via the detection of their strong-decay daughters. In order to check the presence of a suppression in the production of the resonances and to study whether the strength of the suppression is related to the system size, the ratios of the \pT-integrated particle yields \simplerho/\pion, \simplekstarZ/\K, \simplesigmastar/\rmSigma, \simplelambdastar/\rmLambda, \simplexistar/\rmXi~and  \rmphi/\K~have been reported as a function of the cubic root of the charged particle multiplicity density $\langle$\dnchdeta $\rangle^{1/3}$ for different colliding systems; pp at 7~TeV, \pPb~at 5.02~TeV, \PbPb~at  2.76 and 5.02~TeV and \XeXe~collisions at 5.44~TeV nucleon-nucleon centre of mass energies (Fig.~\ref{fig:ratios}). 
The lifetimes of the resonances reported in this figure increase from the top down (1.3~\fmc~(\rmrho)~$<$~4.2~\fmc~(\kstarZ)~$<$~5.5~\fmc~(\Sigmastar)~$<$~12.6~\fmc~(\Lambdastar) $<$~21.7~\fmc~(\Xistar)~$<$~46.4~\fmc~(\phir)). 
A centrality-dependent suppression is clearly observed for \simplerho/\pion, \simplekstarZ/\K~and \simplelambdastar/\rmLambda~in \PbPb~collisions. 
The observed suppression may indicate the dominance of re-scattering mechanisms, compared to the regeneration ones. No centrality dependence across the different systems  is observed for resonances such as the \rmphi~and the 
\simplexistar~which live longer than the \simplekstarZ~and the \simplerho. They decay predominantly after the end of the hadronic phase and their yield should not be affected by the regeneration and the re-scattering effect. The behaviour of the \simplesigmastar/\rmLambda~is peculiar considering that \simplesigmastar~and \simplekstarZ~have similar lifetime. The distribution of this ratio is flat in small system (\dAu~and \pPb)~\cite{Sigmastar_ppb} and similar values are measured at RHIC and LHC energies. Furthemore, no suppression is observed in central \AuAu~collisions at  RHIC energies~\cite{STAR_Sigmastar}. The constant behaviour of the yield ratios of excited to ground-state hyperons with same strangeness content has to be confirmed at LHC energies in ion-ion collisions and should indicate that neither regeneration nor re-scattering dominates with increasing collision system size.   
It is interesting to note that the behaviour of the ratios is at least qualitatively reproduced by calculations using the EPOS3 model~\cite{EPOS}, which takes the regeneration and re-scattering effects of the resonance decay particles in the hadronic phase 
explicitly into account by UrQMD~\cite{UrQMD} (see continous curves in Fig.~\ref{fig:ratios}). The same model EPOS3 without an afterburner, i.e. with URQMD off, (dashed cureves in Fig.~\ref{fig:ratios}) is not able to reproduce the observed trend.

Furthermore, a hint of multiplicity-dependent suppression of the \simplekstarZ/\K~and \simplerho/\pion~ratios has been observed in pp and \pPb~collisions, which may be an indication of the presence of a hadron-gas phase in high-multiplicity pp and \pPb~collisions.

%The ratios of \pT-integrated particle yields K∗0/K and phi/K are shown in Fig. 5 as functions of  
%dNdeta |eta|<0.5 . 
%Within their uncertainties the ratios in pp collisions at s = 7 and 13 TeV and in \pPb~collisions at \sNN~=~5.02~TeV are consistent for similar values of 
%hdN ch /dηi |η|<0.5 . 
%There is a hint of a decrease in K ∗0 /K with increasing 
%hdN ch /dηi |η|<0.5 
%in all three collision systems. 

\begin{figure*}
\centering
\includegraphics[width=6.9cm]{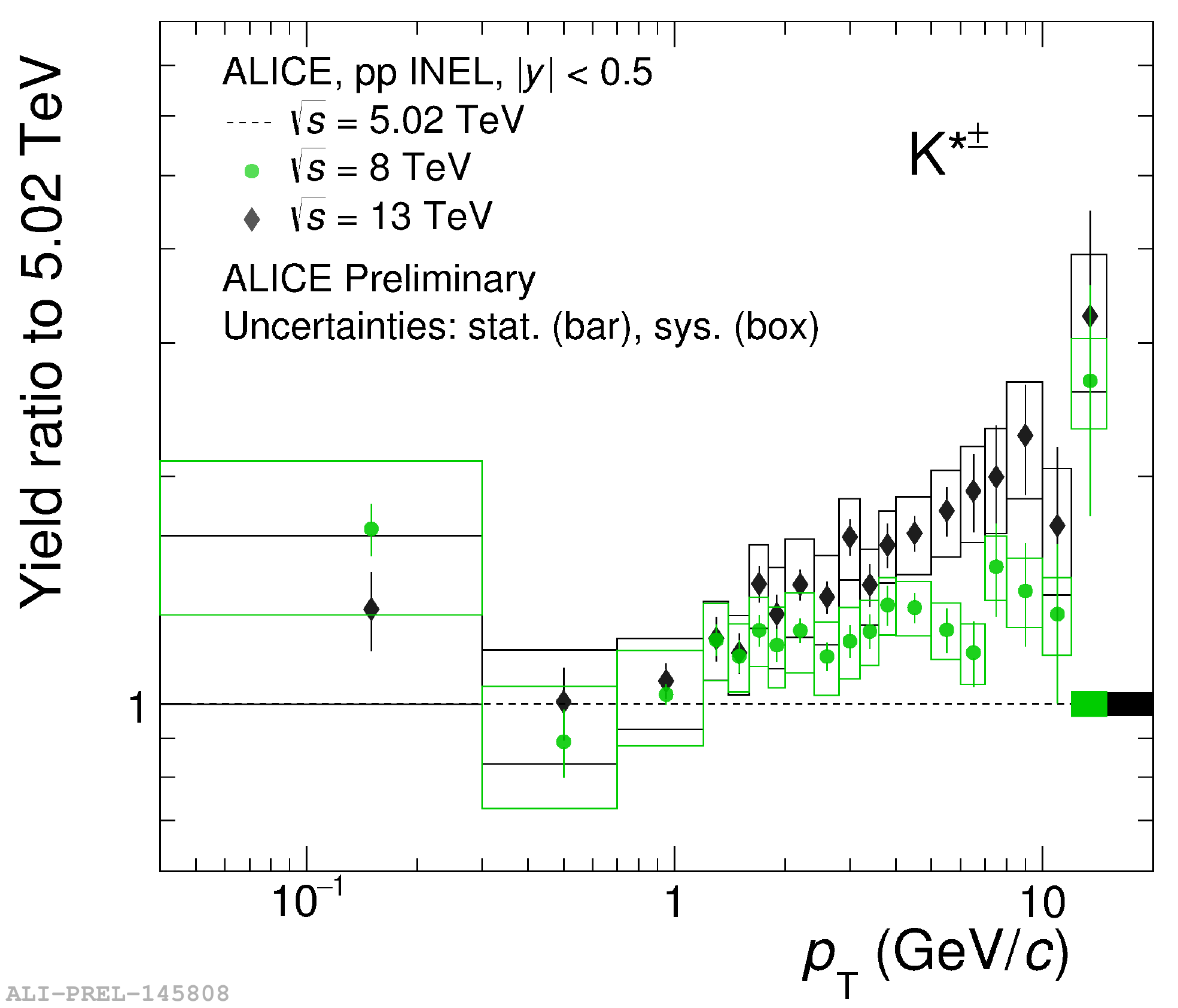}
\includegraphics[width=6.8cm]{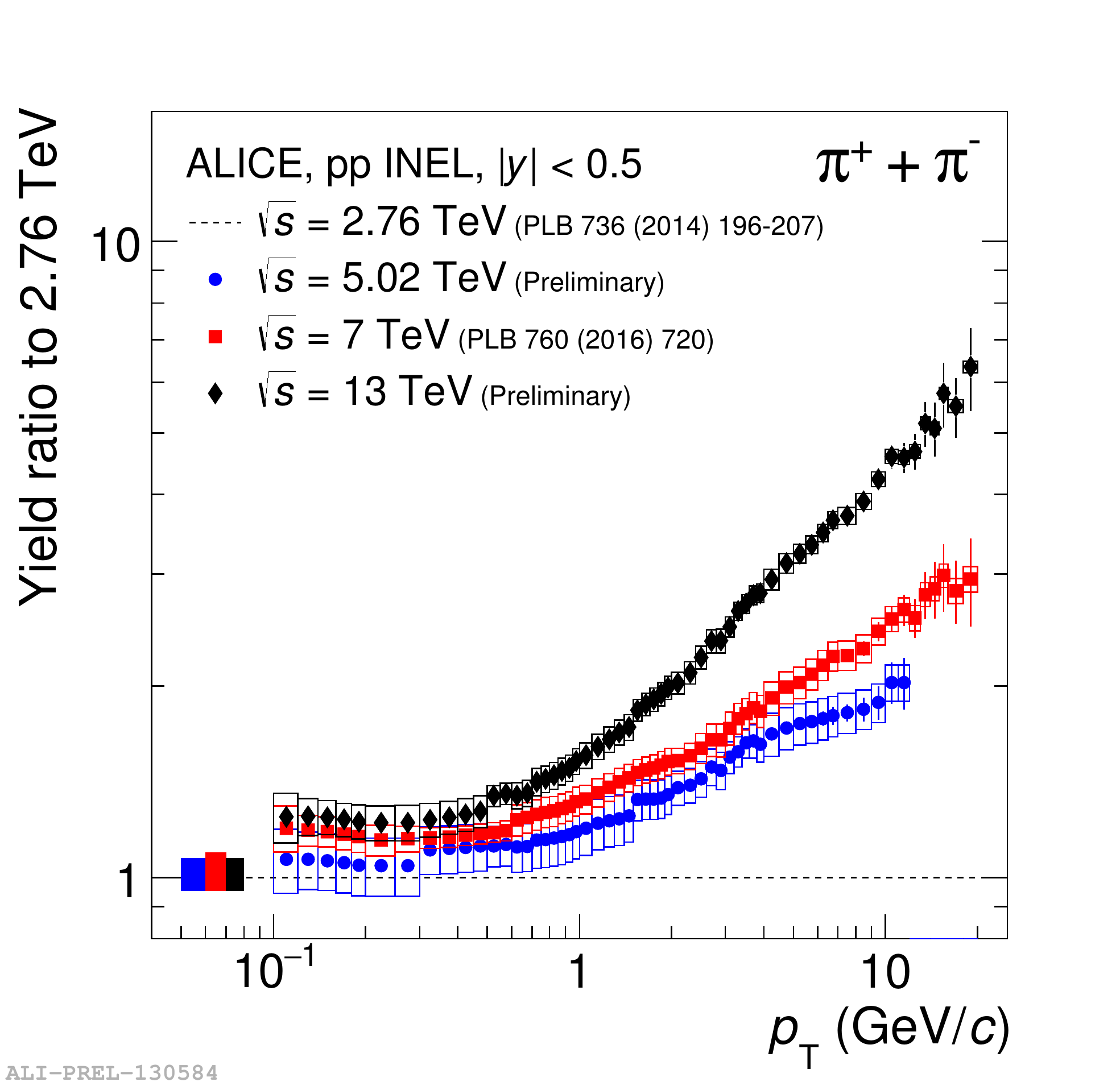}
\caption{(Left panel) Ratios of transverse momentum spectra of \simplekstarch~(left panel) and \pion~(right panel) in inelastic pp events at \s~=~8 and 13~TeV to 5.02~TeV and \s~=~5.02, 7 and 13~TeV to 2.76~TeV, respectively. 
%Statistical and systematic uncertainties are shown by error bars and empty boxes, respectively. Normalization uncertainties are shown as colored boxes around 1 and they are not included in the point-to-point uncertainties.
}
\label{fig:energyratio}       % Give a unique label
\end{figure*}

\begin{figure*}
\centering
\includegraphics[width=7.9cm]{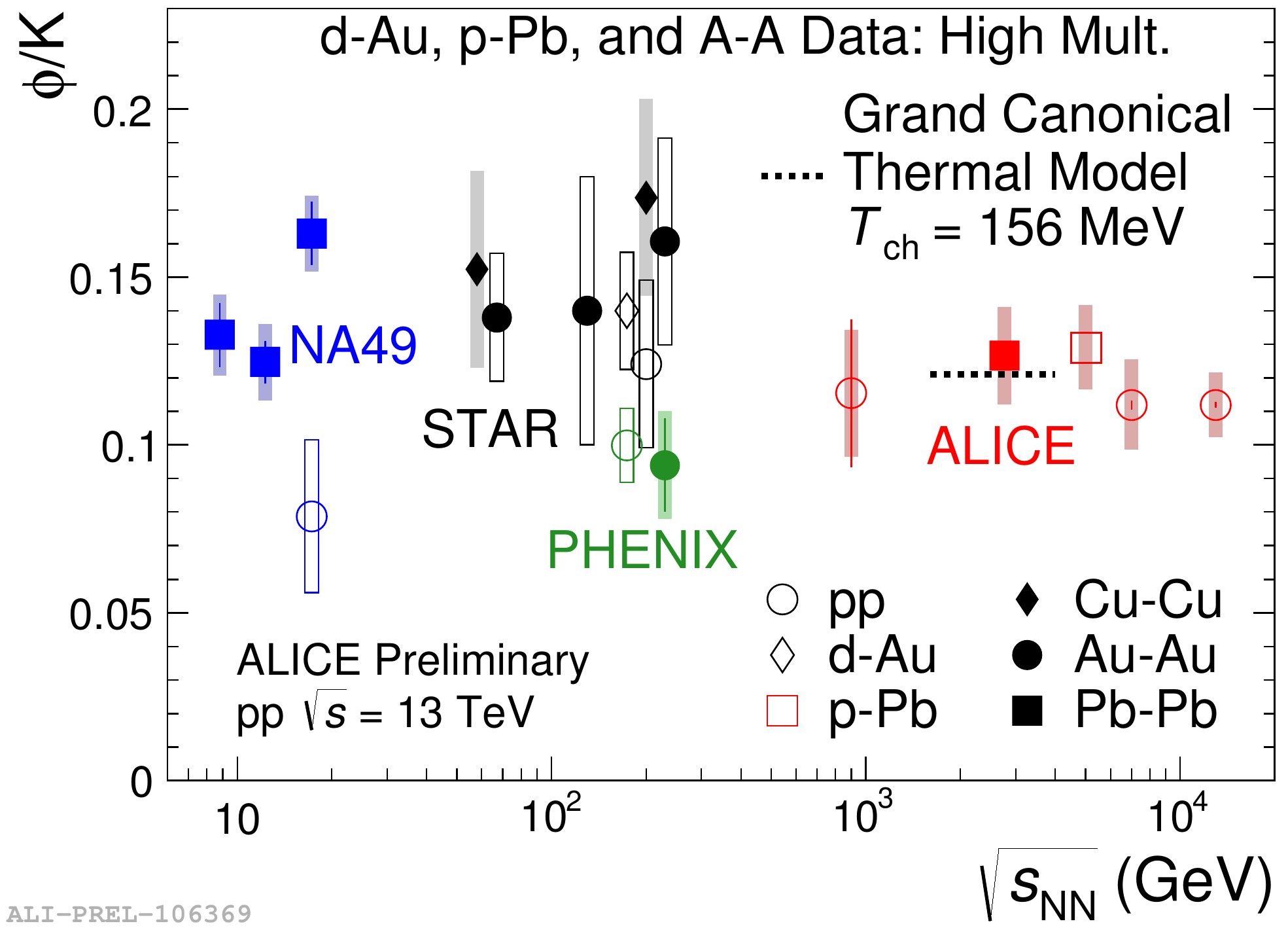}
\includegraphics[width=7.9cm]{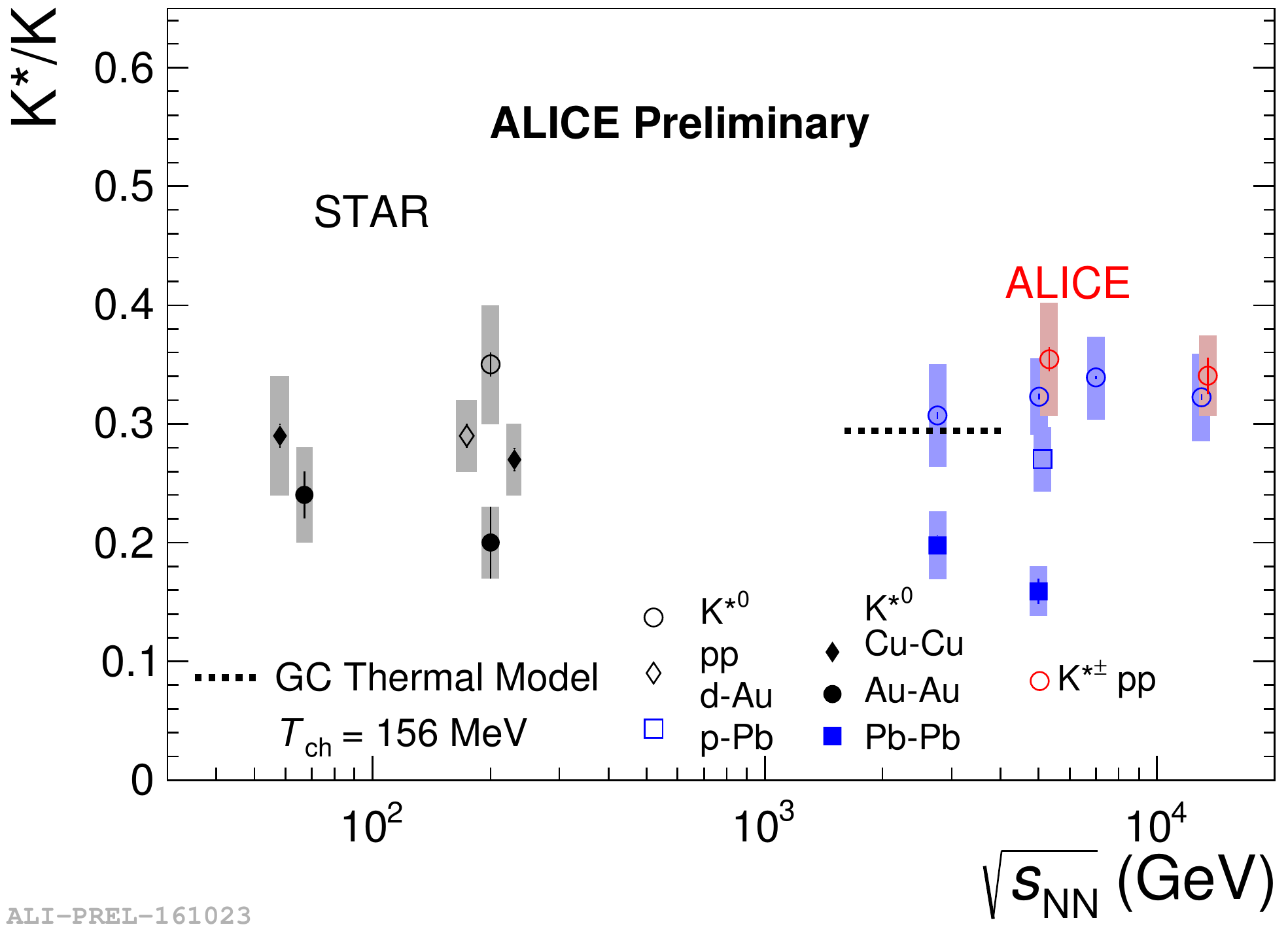}
\caption{Particle ratios \rmphi/\K~(left panel) and \simplekstarch/\K~and \simplekstarZ/\K~(right panel) in 
pp, \dAu, \pPb, \AuAu~and \PbPb~collisions~\cite{phi_kstar_ALICE,STAR_Kstar_200,STAR_Kstar_CuCu_AuAu,STAR_Kstar_dAu,STAR_Kstar_130} 
%{phi_kstar_pp,phi_kstar_ppb,phi_kstar_pbpb,phi_kstar_highpt,STAR_Kstar_200,STAR_Kstar_CuCu_AuAu,STAR_Kstar_dAu,STAR_Kstar_130} 
as a function of the collision enegy \sNN. Some of the data have been shifted horizontally for visibility. The expectations from a thermal model calculation with a chemical freeze-out temperature of 156 MeV~\cite{Stachel_2014} for the most central collisions \PbPb~collisions at \sNN~=~2.76~TeV  are also shown.}
\label{fig:ratiovsenergy}       % Give a unique label
\end{figure*}

\section{Energy dependence of resonance production in pp collisions}

The evolution of the transverse momentum spectra with collision energy is clearly seen in Fig.~\ref{fig:energyratio}. The left and right panels of this figure show the ratios of the \kstarch~transverse-momentum spectra obtained in pp collisions at \s~$=$~8 and 13 TeV to \s~$=$~5.02 TeV and the ratios of \pion~transverse momentum spectra measured in pp collisions at \s~$=$~5.02, 7 and 13 TeV to \s~$=$~2.76~TeV. Both resonance and stable hadrons exhibit the same beahviour. For \pT~$>$~1~\gmom~a clear hardening of the \pT~spectrum is observed when increasing the energy, while at low \pT~the same yield is measured, within the estimated uncertainties, at the studied energies. This suggests that the particle production mechanism in the soft region is independent of the collision energy, while the increase of the slope for \pT~$>$~1~\gmom~suggests an increasing contribution of hard scattering processes in particle production with the collision energy. Similar behaviour has been also observed for other resonances and stable hadrons~\cite{Spectra}.

The  \kstar/\K~and \rmphi/\K~yield ratios (see Fig.~\ref{fig:ratiovsenergy}) do not show a strong dependence on the colliding system or 
the center of mass system energy, and have values equal to the ratios estimated in thermal model calculations with a chemical freeze-out temperature of 156~MeV~\cite{Stachel_2014}. Only \simplekstarZ/\K~ratios measured in central {\mbox{A--A}} collisions both at RHIC and LHC energies are lower, which should be the result of re-scattering and regeneration effects, with the first dominating over the second.

\section{Conclusions}

The latest measurements on resonance production performed for various systems (from pp to \PbPb~including \XeXe)  are reported here. One can observe that particle production is independent of collision system and collision energy at LHC energies and it is driven by the event multiplicity. Moreover a characteristic behaviour  of heavy-ion collision spectra attributed to collective expansion, i.e. a hardening of the \pT~spectra with increasing multiplicity, is observed in elementary collisions also. The origin of this collective flow-like effect in small collision systems is not established. Short-lived resonances as \rmrho, \kstarZ and \Lambdastar~are suppressed in the most central collisions compared to small collision systems, while no suppression is present for resonances such as \Xistar~and \phir~with a lifetime larger than the fireball one at these energies (abot 10 \fmc). The trend of the resonance over stable particle ratios  is
qualitatively described by the EPOS3 model with UrQMD as afterburner to take into account hadronic interactions. An intriguing hint of suppression for short-lived resonances is also observed in high multiplicity pp and \pPb~collisions, suggesting the presence of a hadronic phase also in these elementary collisions. In pp collisions a hardening of the \pT~spectrum is observed when increasing the energy while the relative particle abundance is rather independent of the c.m. system  energy.


\section*{References}
\begin{thebibliography}{9}
%\item Strite S and Morkoc H 1992 {\it J. Vac. Sci. Technol.} B {\bf 10} 1237
\bibitem{Bleicher} Bleicher~M. and Stocker~H.~2004 {\it J. Phys.} G {\bf 46730} S111.
\bibitem{Torrieri} Torrieri~G. and Rafelski~J.~2001 {\it Phys. Lett.} B {\bf 509} 239.
\bibitem{Markert} Markert~C., Bellwied~R., and Vitev~I.~2008 {\it Phys. Lett.} B {\bf 669} 92.
\bibitem{Vogel} Vogel~S., Aichelin~J. and Bleicher~M.~2010 {\it J. Phys.} G {\bf 37} 094046. 
%\bibitem{Bleicher2002} Bleicher~M. and Aichelin~J.~2002 {\it Phys. Lett.} B {\bf 530} 81.
\bibitem{PYTHIA8} Sj\"{o}strand~T., Mrenna~S. and Skands~P.Z~2008 {\it Comput. Phys. Commun.} {\bf 178} 852.
\bibitem{EPOS_LHC} Pierog~T.~\textit{et al.}~2015 {\it Phys. Rev.} C {\bf 92} 034906.
\bibitem{Strangeness_pPb} Adam~J.~\textit{et al.} (ALICE Coll.)~2016 {\it Phys. Lett.} B {\bf 758} 389.
\bibitem{Strangeness_pp} Adam~J.~\textit{et al.} (ALICE Coll.)~2017 {\it Nature Phys.} {\bf 13} 535.
\bibitem{Long_range_CMS} Chatrchyan~S.~\textit{et al.} (CMS Coll.)~2013 {\it Phys. Lett.}  B {\bf 718}  795.
\bibitem{Long_range_ALICE} Abelev~B.~\textit{et al.} (ALICE Coll.)~2013 {\it Phys. Lett.} B {\bf 719} 29.
\bibitem{V2_pp} Acharya~S.~\textit{et al.} (ALICE Coll.)~2019 arXiv:1903.01790.385.
\bibitem{Multiplicity_dep_pp7TeV} Acharya~S.~\textit{et al.} (ALICE Coll.)~2019 {\it Phys. Rev.} C {\bf 99} 024906.
\bibitem{Multiplicity_dep_pPb} Abelev~B.~\textit{et al.} (ALICE Coll.)~2014 {\it Phys. Lett.} B {\bf 728} 25.
\bibitem{Rho_pbpb} Acharya~S.~\textit{et al.} (ALICE Coll.)~2019 {\it Phys. Rev.} C {\bf 99} 064901.
%\bibitem{phi_kstar_pp} Abelev~B.~\textit{et al.} (ALICE Coll.)~2012 {\it Eur. Phys. J.} C {\bf 72} 2183.
%\bibitem{phi_kstar_ppb} Adam~J.~\textit{et al.} (ALICE Coll.)~2016 {\it Eur. Phys. J.} C {\bf 76} 245.
%\bibitem{phi_kstar_pbpb} Abelev~B.~\textit{et al.} (ALICE Coll.)~2015 {\it Phys. Rev.} C {\bf 91} 024609.
%\bibitem{phi_kstar_highpt} Adam~J.~\textit{et al.} (ALICE Coll.)~2017 {\it Phys. Rev.} C {\bf 95} 064606.
\bibitem{phi_kstar_ALICE} Abelev~B.~\textit{et al.} (ALICE Coll.)~2012 {\it Eur. Phys. J.} C {\bf 72} 2183; Adam~J.~\textit{et al.} (ALICE Coll.)~2016 {\it Eur. Phys. J.} C {\bf 76} 245;Abelev~B.~\textit{et al.} (ALICE Coll.)~2015 {\it Phys. Rev.} C {\bf 91} 024609;
Adam~J.~\textit{et al.} (ALICE Coll.)~2017 {\it Phys. Rev.} C {\bf 95} 064606.
\bibitem{Sigmastar_pp} Abelev~B.~\textit{et al.} (ALICE Coll.)~2015 {\it Eur. Phys. J.} C {\bf 75} 1.
\bibitem{Sigmastar_ppb} Adamová~D.~\textit{et al.} (ALICE Coll.)~2017 {\it Eur. Phys. J.} C {\bf 77} 379.
\bibitem{Lambdastar_pbpb} Acharya~S.~\textit{et al.} (ALICE Coll.)~2019 {\it Phys. Rev.} C {\bf 99} 024905; Acharya~S.~\textit{et al.} (ALICE Coll.)~2019  arXiv:1909.00486.
\bibitem{alice} Aamodt~K.~\textit{et al.} (ALICE Coll.)~2008 {\it J. Instrum.} {\bf 3} S08002.
\bibitem{alice_perf} Abelev~B.~\textit{et al.} (ALICE Coll.)~2014 {\it Int. J. Mod. Phys.} A {\bf 29} 1430044.
\bibitem{DIPSY} Bierlich~C., Gustafson~G., Lönnblad~L. and Tarasov~A.~2015 {\it JHEP} {\bf 03} 148.
\bibitem{Ganoti_spectra} Ganoti~P., Proc. of the HNPS2018, the 27th Annual Symposium of the Hellenic Nuclear Physics Society;
 8-9 June 2018, (Athens); \textit{https://eproceedings.epublishing.ekt.gr/index.php/hnps/issue/view/84/showToc}.
\bibitem{Multiplicity_dep_PbPb} Abelev~B.~\textit{et al.} (ALICE Coll.)~2013 {\it Phys. Rev.} C {\bf 88} 044910.
%%%%\bibitem{phi900} B.~Abelev \textit{et al.} (ALICE Coll.), Eur. Phys. J. C \textbf{71}, 1594 (2011)
\bibitem{STAR_Sigmastar} Abelev~B.I.~\textit{et al.} (STAR Coll.)~2006 {\it Phys. Rev. Lett.} {\bf 97} 132301.
\bibitem{EPOS} Knospe~A.~G.~\textit{et al.}~2016 {\it Phys. Rev.} C {\bf 93} 014911.
\bibitem{UrQMD} Bass~S.~A.~\textit{et al.}~1998 {\it Prog. Part. Nucl. Phys.} {\bf 41} 255; Bleicher~M.~\textit{et al.}~1999 {\it J. Phys.} G {\bf 25} 1859.
\bibitem{STAR_Kstar_200} Adams~J.~\textit{et al.} (STAR Coll.)~2005 {\it Phys. Rev.} C {\bf 71} 064902.
\bibitem{STAR_Kstar_CuCu_AuAu} Aggarwal~M.M.~\textit{et al.} (STAR Coll.)~2011 {\it Phys. Rev.} C {\bf 84} 034909.
\bibitem{STAR_Kstar_dAu} Abelev~B.I.~\textit{et al.} (STAR Coll.)~2008 {\it Phys. Rev.} C {\bf 78} 044906.
\bibitem{STAR_Kstar_130} Adler~C.~\textit{et al.} (STAR Coll.)~2002 {\it Phys. Rev.} C {\bf 66} 061901.
\bibitem{Spectra} Adam~J.~\textit{et al.} (ALICE Coll.)~2016 {\it Phys. Lett.} B {\bf 753} 319; Bencedi~G. Proc. EPS-HEP 2017, Eur. Phys. Society conf. on High Energy Physics, 5-12 July 2017 (Venice, Italy); PoS EPS-HEP2017 (2018) 359.
\bibitem{Stachel_2014} Stachel~J.~\textit{et al.}~2014 {\it J. Phys. Conf.} {\bf 509} 012019. 
%\bibitem{iopartnum} IOP Publishing is to grateful Mark A Caprio, Center for Theoretical Physics, Yale University, for permission to include the {\tt iopart-num} \BibTeX package (version 2.0, December 21, 2006) with  this documentation. Updates and new releases of {\tt iopart-num} can be found on \verb"www.ctan.org" (CTAN). 
\end{thebibliography}
\end{document}